\documentclass[aps,prd,reprint,superscriptaddress,nofootinbib]{revtex4-2}
\usepackage[english]{babel}
\usepackage{graphicx} 
\usepackage{microtype}
\usepackage{amssymb}
\usepackage{amsmath}
\usepackage{tabularx}
\usepackage{multirow}
\usepackage{cancel}
\usepackage{lipsum}
\usepackage[dvipsnames]{xcolor}
\RequirePackage[colorlinks=true
,urlcolor=blue
,anchorcolor=blue
,citecolor=blue
,filecolor=blue
,linkcolor=blue
,menucolor=blue
,linktocpage=true
,pdfa=true
]{hyperref}

\usepackage{pstricks}
\usepackage{pstricks-add}
\usepackage{orcidlink}

\newgray{lightgray}{0.90}\newgray{lg1}{0.96}
\newgray{lg2}{0.90}
\newgray{lg3}{0.84}
\newgray{lg4}{0.78}

\date{September 2026}

\begin{document}

\title{Impact of the position effects in the energy calibration of the BULLKID detector}

\newcommand{\AffI}{Dipartimento di Fisica, Sapienza Università di Roma, P. le A. Moro 2, 00185 Roma, Italy}
\newcommand{\AffII}{INFN Sezione di Roma, P.le A. Moro 2, 00185 Roma, Italy}
\newcommand{\AffIII}{Dipartimento di Fisica "Enrico Fermi", Università di Pisa, Largo Bruno Pontecorvo 3, 56127 Pisa, Italy}
\newcommand{\AffIV}{INFN Sezione di Pisa, Largo Bruno Pontecorvo 3, 56127 Pisa, Italy}
\newcommand{\AffV}{Instituto de Física, Universidad Nacional Autónoma de México, A.P. 20-364, Ciudad de México 01000, México}
\newcommand{\AffVI}{INFN Sezione di Ferrara, Via Saragat 1, 44122 Ferrara, Italy}
\newcommand{\AffVII}{Université Grenoble Alpes, CNRS, Grenoble INP, Institut Néel, 38000 Grenoble, France}
\newcommand{\AffVIII}{INFN Laboratori Nazionali del Gran Sasso, 67100 Assergi (AQ) - Italy}
\newcommand{\AffIX}{Institute for Data Processing and Electronics, Karlsruhe Institute of Technology, Hermann-von-Helmholtz-Platz 1 76344, Eggenstein-Leopoldshafen - Germany}
\newcommand{\AffX}{INFN - TIFPA, Via Sommarive 14, 38123 Povo (Trento) Italy}
\newcommand{\AffXI}{Dipartimento di Fisica e Scienze della Terra, Università di Ferrara, Via Saragat 1, 44100 Ferrara - Italy}
\newcommand{\AffXIII}{Gran Sasso Science Institute, Viale F. Crispi, 7 67100 L'Aquila}
\newcommand{\AffXIV}{Now at: Università di Roma "Tor Vergata", Dipartimento di Fisica, Via della Ricerca Scientifica 1, 00133 Roma, Italy}

\author{M.~Folcarelli\orcidlink{0009-0009-7799-2515}}\email{matteo.folcarelli@uniroma1.it}\affiliation{\AffI}\affiliation{\AffII}
\author{M.~Cappelli\orcidlink{0009-0002-6148-5964}}\email{matteo.cappelli@uniroma1.it}\affiliation{\AffI}\affiliation{\AffII}
\author{G.~Del~Castello\orcidlink{0000-0001-7182-358X}}\affiliation{\AffII}
\author{C.~Bonomo\orcidlink{0009-0001-0336-3857}}\affiliation{\AffI}\affiliation{\AffII}
\author{A.~Cruciani\orcidlink{0000-0003-2247-8067}}\affiliation{\AffII}
\author{D.~Delicato\orcidlink{0009-0005-0516-6872}}\affiliation{\AffI}\affiliation{\AffII}\affiliation{\AffVII}
\author{L.~Pesce\orcidlink{0009-0001-5659-4691}}\affiliation{\AffI}\affiliation{\AffII}
\author{D.~Quaranta\orcidlink{0009-0000-2954-4456}}\affiliation{\AffI}\affiliation{\AffII}
\author{M.~Vignati\orcidlink{0000-0002-8945-1128}}\affiliation{\AffI}\affiliation{\AffII}

\begin{abstract}
    BULLKID is a cryogenic solid-state detector based on an array of silicon dice equipped with Kinetic Inductance Detectors (KIDs) for low-mass dark matter searches and coherent elastic neutrino-nucleus scattering. Energy calibration can be performed using optical LED photon bursts, which are absorbed within the first few hundred nanometers of the substrate. An energy calibration using a $^{241}$Am gamma-ray source revealed a $<10\%$ discrepancy with the LED calibration. A recent study by Matava and Williams suggested that phonon statistics and phonon collection varying with the interaction position could explain such a discrepancy. This work evaluates the impact of these effects on the calibration of the BULLKID detector, demonstrating they both have a negligible effect. We clarify the results obtained during the $^{241}$Am calibration regarding the origin of the observed discrepancy.
\end{abstract}

\maketitle

\section{Introduction}
BULLKID \cite{Cruciani} is a cryogenic solid-state detector designed for direct searches of particle Dark Matter candidates, with mass of order $\text{GeV}/\text{c}^2$ or below, and coherent elastic neutrino-nucleus scattering. It is based on an array of dice carved in a $5~$mm-thick silicon crystal, sensed by phonon-mediated aluminum Kinetic Inductance Detectors (KIDs). The energy calibration can be performed with bursts of optical photons \cite{Cruciani, Lantern}, a flexible technique able to calibrate the device in a range spanning from a few eVs to tens of keVs. When a burst of photons emitted from an LED interacts in a silicon crystal, it produces phonons. Part of them reach the sensor, break Cooper pairs in the superconductor and generate a signal. 
By leveraging the intrinsic statistics of the photons absorbed by the silicon, the energy calibration function can be extracted.

Optical photons are absorbed in the first hundreds of nanometers of the die and may not generate exactly the same signal as particles interacting in the bulk. To validate this calibration procedure against bulk events, the BULLKID collaboration conducted a dedicated measurement with the use of a $^{241}$Am source~\cite{Americium} that emits $59.5~$keV $\gamma$-rays. The absorption depth in silicon is $\lambda_{\text{Am}}=13~$mm~\cite{XCOM}, resulting in a quite uniform interaction probability across the entire die. The measurement proved that the reconstructed energy of the photo-peak, calibrated using the LED, presents a $<10\%$ deficit with the nominal energy, validating the optical calibration to such a level of accuracy, in line with the requirement of present experiments. 

The segmented structure of the detector was exploited to study the position effects in the case of bulk events. The phonons generated from a particle interaction in a die partially leak to neighboring dice through the $0.5~$mm-thick common disk holding the array. As a consequence, to any particle signal detected by a KID, corresponds a coincident and smaller signal in all its neighbors. We observed that the reconstructed amplitude of the Americium photo-peak evaluated by a KID is anti-correlated with the amplitude in coincidence with all its neighbors. We motivated such an observation in Ref.~\cite{Americium} in terms of a position dependent phonon leakage, that is not expected in the case of LED photons. In fact, they present a smaller absorption length $\lambda_{\text{LED}}=180~$nm, and the typical dimension of the light spot on the die is approximately $1~$mm, resulting in interactions much more localized in space than for Americium $\gamma$-rays. 

A recent work from W.~Matava and M.R.~Williams~\cite{matava} developed a formalism for the optical calibration in which, in addition to the Poisson statistics of photons, other effects are taken into account. Among these, a possible variation of the phonon collection efficiency, due to the different impact points of the LED photons, may have a non-negligible influence in some circumstances. Such a variation of the phonon collection is mentioned in the paper as the most-likely cause of the mismatch between the two calibrations in BULLKID. It is consequently suggested to modify the model behind the optical calibration accordingly.

In this work we evaluate the impact of the effects pointed out by Matava and Williams in Ref.~\cite{matava} on BULLKID and we conclude that they are negligible. We therefore clarify the result of Ref.~\cite{Americium} regarding the measured difference between surface (LED) and bulk ($^{241}$Am) events. The document is organized as follows: Section~\ref{sec:optical_calibration} reports a brief summary of the results presented in Ref.~\cite{matava}, restricted to the effect of a variable phonon collection, and not considering effects such as the Fano Factor or the phonon energy distribution, since they are already recognized as subdominant. Section~\ref{sec:bullkid_case} illustrates the application to the BULLKID optical calibration.

\section{Optical calibration}
\label{sec:optical_calibration}
The reconstructed amplitude $\mathcal{I}$ of an LED pulse in a detector is:
\begin{equation}
    \mathcal{I}=r\epsilon_{ph}N^\dagger_{ph}~,
\end{equation}
where $r~$ is a scale factor converting energy to an arbitrary unit of the amplitude, $\epsilon_{ph}$ is the mean energy of phonons and $N^\dagger_{ph}$ is the number of phonons detected by the sensor. The mean value of the reconstructed amplitude is then:
\begin{equation}
  E[\mathcal{I}] =r\epsilon_{ph}E[N^\dagger_{ph}]=rE_\gamma\lambda\mu_Q=I_0\cdot\lambda~,
\end{equation}
where we used the result $E[N^\dagger_{ph}]=\lambda\mu_QE_\gamma/\epsilon_\text{ph}$ obtained in Ref.~\cite{matava}. Specifically, $E_\gamma$ is the photon energy, $\lambda$ is the mean number of emitted photons per burst and $\mu_Q$ is the mean phonon collection efficiency in the case of LED events. Additionally, $I_0=rE_\gamma\mu_Q$ is defined as the mean reconstructed amplitude for a single photon absorption. The responsivity $R$ from a known deposited energy to amplitude is then 
\begin{equation}
    R=I_0/E_\gamma=r\cdot \mu_Q~.
    \label{eq:resp}
\end{equation}

The variance of the amplitude distribution is evaluated by accounting both for the variance of the number of phonons and for the baseline resolution of the device that is supposed to be distributed as a gaussian $\mathcal{N}(0,\sigma_i^2)$:
\begin{align}
\text{VAR}[\mathcal{I}] &=r^2\epsilon^2_{ph}\text{VAR}[N^\dagger_{ph}]+\sigma^2_i\nonumber\\
&=r^2\epsilon_{ph}^2(a^2\cdot\lambda\cdot(\sigma_Q^2+\mu_Q^2)+a\lambda\mu_Q)+\sigma^2_i\nonumber\\
&=E[\mathcal{I}]~I_0\cdot(1+\frac{\sigma^2_Q}{\mu^2_Q}+\frac{1}{\mu_Q~a})+\sigma^2_i\nonumber\\
&=E[\mathcal{I}]~I_0\cdot(1+\delta)+\sigma^2_i~,
\end{align}
with
\begin{equation}
  \delta = \frac{\sigma^2_Q}{\mu^2_Q}+\frac{1}{\mu_Q~a} = \frac{\sigma^2_Q}{\mu^2_Q}+\frac{\epsilon_{ph}}{\mu_Q~E_\gamma}~,
  \label{eq:delta}
\end{equation}
where $\text{VAR}[N^\dagger_{ph}]$ is derived by Matava and Williams in Ref.~\cite{matava}, $\sigma^2_Q$ is the variance of the phonon collection efficiency distribution and $a=E_\gamma/\epsilon_\text{ph}$.

The standard optical calibration neglects $\delta$ and infers the responsivity of the detector $R$ by fitting the slope of the $E[\mathcal{I}]$ vs. $\text{VAR}[\mathcal{I}]$ curve, knowing $E_\gamma$. 

Generally $\delta$ can be neglected when comparable to or smaller than the required precision of the calibration. Both terms composing it should be estimated and taken into account only if necessary:
\begin{itemize}
    \item the $\frac{1}{\mu_Qa}$ term coming from the phonon statistics, which appears when including the Poisson distribution of the generated phonons, after a single photon absorption in the substrate. It must be taken into account when $a\cdot \mu_Q\sim 1$.
    \item the $\sigma_Q^2/\mu^2_Q$ term accounting for effects of the position of interactions on the phonon collection efficiency. 
\end{itemize}

\section{BULLKID case}
\label{sec:bullkid_case}
The $\delta$ term in Eq.~\ref{eq:delta} is not accounted for in the LED calibration of BULLKID. Here, we verify if it could be the source of the difference from the $\gamma$-ray calibration in Ref.~\cite{Americium}.

The phonon collection efficiency $Q$ corresponds to the total phonon to quasiparticle conversion efficiency $\eta$ derived in Ref.~\cite{Cruciani}\footnote{The BULLKID collaboration usually refers to the phonon collection efficiency as $\eta/\eta_0\sim 40\%$, but we think that the definition of $Q$ refers to $\eta$ in this case} from the optical calibration. Therefore, we take as its mean value
\begin{equation}
    \mu_Q=\eta = 24\%~.
\end{equation}
BULLKID adopts a $400~$nm LED ($E_\gamma = 3.1$ eV), so the mean number of phonons generated from a single photon $a$ is:
\begin{equation}
    a = E_\gamma/\epsilon_{ph}= (3.1~\text{eV})/(0.34~\text{meV})= 9.1\cdot10^3~,
\end{equation}
where we assumed, as also done in Ref.~\cite{matava}, that the mean phonon energy is twice the superconductive band of aluminum $\epsilon_{ph}=2\Delta_\text{Al}=0.34~$meV\footnote{We agree that this can be an underestimation of the mean phonon energy distribution}.
The contribution of the phonon statistics to $\delta$ is then:
\begin{equation}
    \frac{1}{\mu_Qa} = 4.6\cdot 10^{-4}\ll1~.
\end{equation}
As already concluded in Ref.~\cite{matava}, we confirm that this term can be excluded as a possible source of mismatch mentioned in Ref.~\cite{Americium}.

In Ref.~\cite{Americium} we observed an anti-correlation between the amplitude reconstructed by a KID and the amplitude in coincidence with its neighbors.
We model this effect as originating from a phonon collection varying along the BULLKID die. Depending on the interaction point, the more phonons are collected by the KID on the interaction die, the fewer can leak into its neighbors, hence producing the observed anti-correlation. The variation of $Q$ can be constrained from the measured width of the Americium photo-peak of $5\%$ as 
\begin{equation}
    \sigma_Q < 5\%~\mu_Q~,
\end{equation} 
for events distributed inside the entire die. Even assuming such a variation of $Q$ for the LED photons, which interact in a very localized region of the surface of the die, with a 1 mm lateral dimension at a distance of 5 mm from the sensor, the contribution to $\delta$ is:
\begin{equation}
    \frac{\sigma^2_Q}{\mu^2_Q}<(5\%)^2=2.5\cdot 10^{-3}\ll1~.
\end{equation}
Therefore, this contribution is also negligible and we conclude that overall the $\delta$ term in Eq.~\ref{eq:delta} does not explain the disagreement between LED and $\gamma$-ray calibrations.

\begin{figure}[tb]
    \centering
    \includegraphics[width=0.9\linewidth]{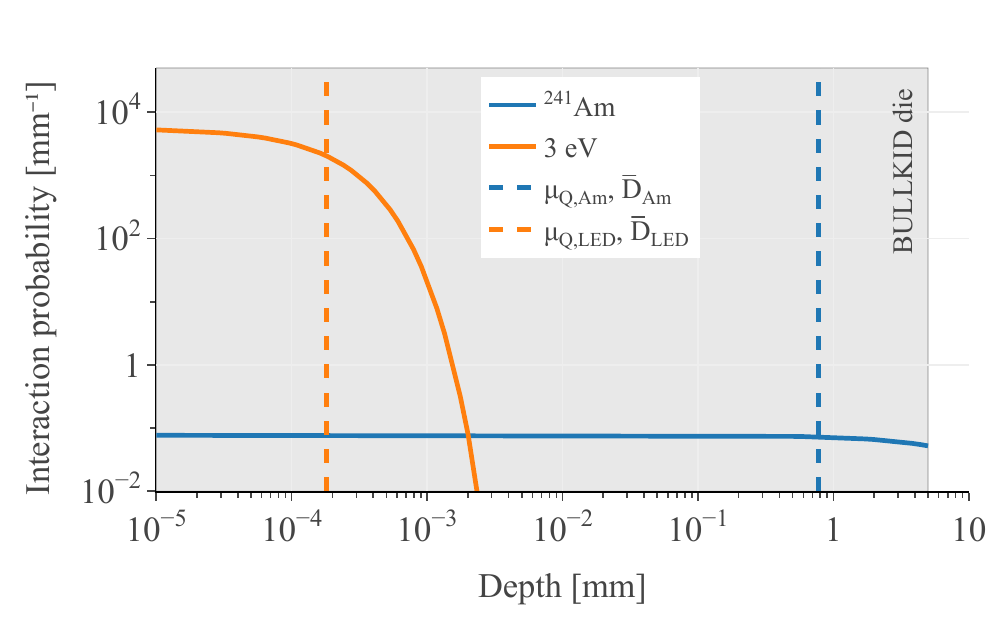}
    \caption{Interaction probability density distribution of the $59.5~$keV $\gamma$-rays emitted by Am and of the $3~$eV optical photons as a function of the absorption depth. The dimension of the silicon die of BULLKID is indicated. The mean of the two distributions is different and indicated with the vertical dashed lines. Different mean values $\mu_Q$ of the phonon collection efficiency are obtained for the two interactions.}
    \label{fig:qsampling}
\end{figure}

We clarify the conclusions of Ref.~\cite{Americium} and underline the fact that $Q$ varies depending on the distance between the interaction point and the common disk where phonons leak. Due to the different penetration of $59.5~$keV gammas and optical photons and the subsequent different phonon leakage, different portions of the distribution of $Q$ are sampled by the two calibrations, leading to different $\mu_Q$. Figure~\ref{fig:qsampling} shows the probability density of the optical and Americium photons to be absorbed in silicon as a function of the penetration depth. In the first case, the photons are absorbed in the first hundreds of nanometers of the crystal while in the second case an almost uniform distribution along the entire silicon die can be noticed. The means of the two distributions are indeed very different:
\begin{align*}
    \bar D_\text{LED}&=\int_0^{L}\frac{x}{\lambda_\text{LED}} \frac{e^{-x/\lambda_{\rm LED}}}{1-e^{-L/\lambda_{\rm LED}}}~dx=\lambda_\text{LED}=180~\text{nm}~,\\
    \bar D_\text{Am}&=\int_0^{L}\frac{x}{\lambda_\text{Am}} \frac{e^{-x/\lambda_{\rm Am}}}{1-e^{-L/\lambda_{\rm Am}}}~dx\sim2.34~\text{mm}~,
\end{align*}
leading to different values of $\mu_Q$. $L$ is the thickness of the die.
Since the responsivity $R$ in Eq.~\ref{eq:resp} linearly depends on $\mu_Q$, two different estimates of this parameter are obtained in the two cases of LED and Americium events:
\begin{equation}
    R_\text{LED}=r\cdot \mu_\text{Q,LED}\hspace{0.5cm}R_\text{Am}=r\cdot \mu_\text{Q,Am}~.
\end{equation}

In conclusion, we can point out the different mean value of the phonon collection between surface and bulk events as the most probable cause of the systematics of the optical calibration with respect to the one performed with $\gamma$-rays.

\begin{acknowledgments}
This work was supported by the INFN, Sapienza University of Rome and co-funded by the European Union (ERC, DANAE, Grant No. 101087663). Views and opinions expressed are however those of the author(s) only and do not necessarily reflect those of the European Union or the European Research Council. Neither the European Union nor the granting authority can be held responsible for them. We acknowledge the support of the PTA platform for the fabrication of the device. We thank A. Mazzolari for useful discussions.
\end{acknowledgments}


\begin{thebibliography}{5}%
\makeatletter
\providecommand \@ifxundefined [1]{%
 \@ifx{#1\undefined}
}%
\providecommand \@ifnum [1]{%
 \ifnum #1\expandafter \@firstoftwo
 \else \expandafter \@secondoftwo
 \fi
}%
\providecommand \@ifx [1]{%
 \ifx #1\expandafter \@firstoftwo
 \else \expandafter \@secondoftwo
 \fi
}%
\providecommand \natexlab [1]{#1}%
\providecommand \enquote  [1]{``#1''}%
\providecommand \bibnamefont  [1]{#1}%
\providecommand \bibfnamefont [1]{#1}%
\providecommand \citenamefont [1]{#1}%
\providecommand \href@noop [0]{\@secondoftwo}%
\providecommand \href [0]{\begingroup \@sanitize@url \@href}%
\providecommand \@href[1]{\@@startlink{#1}\@@href}%
\providecommand \@@href[1]{\endgroup#1\@@endlink}%
\providecommand \@sanitize@url [0]{\catcode `\\12\catcode `\$12\catcode
  `\&12\catcode `\#12\catcode `\^12\catcode `\_12\catcode `\%12\relax}%
\providecommand \@@startlink[1]{}%
\providecommand \@@endlink[0]{}%
\providecommand \url  [0]{\begingroup\@sanitize@url \@url }%
\providecommand \@url [1]{\endgroup\@href {#1}{\urlprefix }}%
\providecommand \urlprefix  [0]{URL }%
\providecommand \Eprint [0]{\href }%
\providecommand \doibase [0]{https://doi.org/}%
\providecommand \selectlanguage [0]{\@gobble}%
\providecommand \bibinfo  [0]{\@secondoftwo}%
\providecommand \bibfield  [0]{\@secondoftwo}%
\providecommand \translation [1]{[#1]}%
\providecommand \BibitemOpen [0]{}%
\providecommand \bibitemStop [0]{}%
\providecommand \bibitemNoStop [0]{.\EOS\space}%
\providecommand \EOS [0]{\spacefactor3000\relax}%
\providecommand \BibitemShut  [1]{\csname bibitem#1\endcsname}%
\let\auto@bib@innerbib\@empty
\bibitem [{\citenamefont {Cruciani}\ \emph {et~al.}(2022)\citenamefont
  {Cruciani} \emph {et~al.}}]{Cruciani}%
  \BibitemOpen
  \bibfield  {author} {\bibinfo {author} {\bibfnamefont {A.}~\bibnamefont
  {Cruciani}} \emph {et~al.},\ }\href {https://doi.org/10.1063/5.0128723}
  {\bibfield  {journal} {\bibinfo  {journal} {Applied Physics Letters}\
  }\textbf {\bibinfo {volume} {121}},\ \bibinfo {pages} {213504} (\bibinfo
  {year} {2022})}\BibitemShut {NoStop}%
\bibitem [{\citenamefont {Castello}(2024)}]{Lantern}%
  \BibitemOpen
  \bibfield  {author} {\bibinfo {author} {\bibfnamefont {G.~D.}\ \bibnamefont
  {Castello}},\ }\href
  {https://doi.org/https://doi.org/10.1016/j.nima.2024.169728} {\bibfield
  {journal} {\bibinfo  {journal} {Nuclear Instruments and Methods in Physics
  Research Section A: Accelerators, Spectrometers, Detectors and Associated
  Equipment}\ }\textbf {\bibinfo {volume} {1068}},\ \bibinfo {pages} {169728}
  (\bibinfo {year} {2024})}\BibitemShut {NoStop}%
\bibitem [{\citenamefont {Folcarelli}\ \emph {et~al.}(2026)\citenamefont
  {Folcarelli} \emph {et~al.}}]{Americium}%
  \BibitemOpen
  \bibfield  {author} {\bibinfo {author} {\bibfnamefont {M.}~\bibnamefont
  {Folcarelli}} \emph {et~al.},\ }\href
  {https://doi.org/10.1140/epjc/s10052-026-15523-4} {\bibfield  {journal}
  {\bibinfo  {journal} {The European Physical Journal C}\ }\textbf {\bibinfo
  {volume} {86}},\ \bibinfo {pages} {301} (\bibinfo {year} {2026})}\BibitemShut
  {NoStop}%
\bibitem [{\citenamefont {Berger}\ \emph {et~al.}(1999)\citenamefont {Berger},
  \citenamefont {Hubbell}, \citenamefont {Seltzer}, \citenamefont {Coursey},\
  and\ \citenamefont {Zucker}}]{XCOM}%
  \BibitemOpen
  \bibfield  {author} {\bibinfo {author} {\bibfnamefont {M.}~\bibnamefont
  {Berger}}, \bibinfo {author} {\bibfnamefont {J.}~\bibnamefont {Hubbell}},
  \bibinfo {author} {\bibfnamefont {S.}~\bibnamefont {Seltzer}}, \bibinfo
  {author} {\bibfnamefont {J.}~\bibnamefont {Coursey}},\ and\ \bibinfo {author}
  {\bibfnamefont {D.}~\bibnamefont {Zucker}},\ }\href@noop {} {\bibinfo {title}
  {Xcom: Photon cross section database (version 1.2)}},\ \bibinfo
  {howpublished} {\url{http://physics.nist.gov/xcom}} (\bibinfo {year}
  {1999}),\ \bibinfo {note} {national Institute of Standards and
  Technology}\BibitemShut {NoStop}%
\bibitem [{\citenamefont {Matava}\ and\ \citenamefont
  {Williams}(2026)}]{matava}%
  \BibitemOpen
  \bibfield  {author} {\bibinfo {author} {\bibfnamefont {W.}~\bibnamefont
  {Matava}}\ and\ \bibinfo {author} {\bibfnamefont {M.~R.}\ \bibnamefont
  {Williams}},\ }\href {https://doi.org/10.1103/5vfj-5w47} {\bibfield
  {journal} {\bibinfo  {journal} {Phys. Rev. D}\ }\textbf {\bibinfo {volume}
  {114}},\ \bibinfo {pages} {032007} (\bibinfo {year} {2026})}\BibitemShut
  {NoStop}%
\end{thebibliography}
\end{document}